\documentclass[preprint,12pt]{elsarticle}

\usepackage{amssymb}

\journal{}

\begin{document}

\begin{frontmatter}



\title{Atmospheric Electron Spectrum above 30~GeV at the high altitude}


\author[label1]{Y.~Komori}
\author[label2]{T.~Kobayashi}
\author[label3]{K.~Yoshida}
\author[label4]{J.~Nishimura}
\address[label1]{Kanagawa University of Human Services, Yokosuka 238-8522, Japan, 
	             komori-y@kuhs.ac.jp, 81-46-828-2759}
\address[label2]{Department of Physics, Aoyama Gakuin University, Sagamihara 229-8558, Japan}
\address[label3]{College of Systems Engineering and Science, Shibaura Institute of Technology, Saitama
337-8570, Japan}
\address[label4]{The Institute of Space and Astronautical Science, JAXA, Sagamihara 229-8510, Japan}

\begin{abstract}
We have observed the primary electron spectrum from 30~GeV to 3~TeV
using emulsion chambers flown by balloons at the top of the atmosphere, 
for the purpose of exploring the origin of cosmic rays in the Galaxy. 
The atmospheric gamma rays have been 
simultaneously observed in the 30~GeV $\sim$ 8~TeV energy range.
In this paper, we estimate the atmospheric electron spectrum 
in the upper atmosphere ($<$ 10~${\rm g/cm^2}$) 
from our observed gamma-ray spectrum using the electromagnetic shower theory
in order to derive the primary cosmic-ray electron spectrum.
The transport equations of the electron and gamma-ray spectrum 
are analytically solved and the results are compared with those of Monte Carlo simulation (MC).
Since we used the observed atmospheric gamma rays as the source of atmospheric electrons, 
our solutions are free from ambiguities on the primary cosmic-ray nuclear spectra 
and nuclear interaction models included in MC.
In the energy range above several hundred GeV, 
the Dalitz electrons produced directly from neutral pions contribute
around 10 percent to the whole atmospheric electron spectrum at the depth of 4~${\rm g/cm^2}$, 
which increases in importance at the higher altitude 
and cannot be ignored in TeV electron balloon observations.
%
%
\end{abstract}


\end{frontmatter}


%
\section{Introduction}
%
The primary electrons above 30~GeV using balloon-borne emulsion chambers 
have been observed since 1960's \cite{apj80}\cite{apj04}\cite{apj11}.
The major purpose of primary electron observations is 
to investigate the transport and the origin of cosmic rays in the Galaxy, and 
so far one believes that the most likely sources of electrons are supernova remnants 
such as Vela SNR \cite{apj04}.
High-energy cosmic-ray electrons rapidly lose their energy 
because of the synchrotron radiation in the Galactic magnetic field 
and inverse-Compton scattering with the Galactic photons.
These processes cause a steeper spectrum with the index of $-(3.0 \sim 3.3)$.
On the other hand, cosmic-ray protons have the spectral index of around $-2.7$.
Therefore the atmospheric electrons increase the ratio to primary electrons 
because they have the same spectral index as the protons.
In order to derive the primary electron spectrum in balloon experiments,
it is required to subtract the atmospheric electron spectrum
produced in residual atmosphere.

Atmospheric electrons above 10~GeV are mostly due to the pair productions
of gamma rays produced in the hadronic interactions in the residual atmosphere.
Those gamma rays are mainly due to the decays of $\pi^0$ mesons with smaller fractions
from $\eta$ mesons, $K$ mesons, etc.
In this paper we estimate
the atmospheric electron spectrum from the observed gamma-ray spectrum 
and show the contribution of the Dalitz electrons which 
are directly produced electrons from $\pi^0$ and $\eta$ mesons, 
that is not through gamma rays.
In addition, we consider the direct production of electrons from $K$ and $\pi^{\pm}$ mesons,
which contributes in the lower energy range below 100~GeV 
and increases with decreasing energy.
The first precise estimates of atmospheric electron flux was
calculated by Orth \& Buffington \cite{orth}.
They have given the atmospheric electrons in the 1-100~GeV energy range 
contributed from the decay of mesons produced by nuclear interactions.
%
Their calculations are based on MC techniques and 
they presented just a simple analytic model which summarize their MC results.
On the other hand, we estimate the flux of atmospheric electrons 
from the observed gamma rays by using the electromagnetic shower theory.
The results are compared to the MC results assuming the primary hadron flux 
and nuclear interaction model, 
and discuss the behavior of both electron and gamma-ray spectra 
in the upper atmosphere.
%
\section{Calculations}
We consider the transport equations of 
the electron and gamma-ray spectrum above 10~GeV in the atmosphere ($< 10 {\rm g/cm^2}$)
based on the electromagnetic shower theory of Approximation~A \cite{HB}
which treats the behavior of electromagnetic interaction in the upper atmosphere.
The atmospheric electrons are mostly produced by gamma rays, 
originated from $\pi^0$ decay mode, $\eta$ and $K^0$ decay mode, etc.
The equation should include 
the production rate of atmospheric gamma rays from those mesons.
We estimate the production rate from the atmospheric gamma-ray spectrum  $J_{\gamma}(E)$ 
observed simultaneously in our primary electron experiments \cite{yoshida}:
\begin{eqnarray}
J_{\gamma}(E) {\rm (m^2 \cdot s \cdot sr \cdot GeV)^{-1}} \hspace{35mm} & & \nonumber \\ 
 =  (1.12 \pm 0.13) \times 10^{-4} (E/100{\rm GeV})^{-2.73\pm 0.06} & &
\ ,
\label{ga_ob}
\end{eqnarray}
in the range from 30~GeV to 8~TeV, 
which is the vertical gamma-ray spectrum normalized at 4.0 g/cm$^2$ 
originating from hadronic interactions of primary protons 
and heavy nuclei with the residual atmosphere. 
The contribution from primary electrons is subtracted.
%
\subsection{Formulation and Solution}    \label{sec:sol}
At the depth $t$ in the atmosphere, 
the number of electrons,  $\pi(E, t)dE$, and gamma-rays, $\gamma(E, t)dE$,
with energies between $E$ and $E+dE$ 
satisfy the following simultaneous equations.
\begin{eqnarray}
  \frac{\partial \pi (E, t)}{\partial t} & = & 
   -A' \pi (E, t) + B' \gamma (E, t) +   \pi_{D} (E, t) 
\label{dif1} \\
  \frac{\partial \gamma (E, t)}{\partial t} & = & 
   C' \pi (E, t) - \sigma_{0}\gamma(E, t) + \gamma_{ex} (E, t)
\label{dif2}
\end{eqnarray}
As shown in the literature \cite{HB}, the operator $A'$ 
represents the effect of the radiation loss process, 
$B'$ the pair production, $C'$ the gamma ray contributed from electrons, 
and $\sigma_0$ the absorption by the pair production. 
The functional forms of these operators are shown in \ref{app1}.
%

In eq.~(\ref{dif2}), 
we add the gamma-ray production rate from nuclear interactions as:
\begin{equation}
  \gamma_{ex}(E,t)dE = k_g E^{-\beta-1} \exp(-t/L) dE \ , 
\label{ga_ex}
\end{equation}
in which the spectral index $\beta = 1.73$ is determined from eq.~({\ref{ga_ob}).
The attenuation path length is assumed as $L= 100~{\rm g/cm^2}$ \cite{yoshida},
giving the uncertainty of a few percent to the production rate 
at the depth of 10~${\rm g/cm^2}$.
The coefficient $k_g$ is estimated from the observed gamma ray spectrum of eq.~(\ref{ga_ob}).
Gamma rays are mostly produced through the $\pi^0$ decay mode and 
other small contributions exist from the $\eta$ and $K^0$ decay mode, which give
the fraction of 0.16 and 0.03, respectively, to the $\pi^0$ decay mode 
for the spectral index of around 1.7 
\cite{yoshida}.
%

In the eq.~(\ref{dif1}) we include the small contribution of electrons 
from the nuclear interaction of Dalitz decay mode, which is the second major mode of 
$\pi^0$ decay, producing one real gamma ray and one electron-positron pair, 
occupying 1.2\% of $\pi^0$ decay process \cite{pdb}.
The production rate of Dalitz electron pairs is estimated
as a fraction of the gamma ray production rate as follows:
\begin{equation}
  \pi_{D}(E,t) = a \cdot \gamma_{ex}(E,t) \ .
\label{ratio_a}
\end{equation}
The coefficient $a$ is estimated from the decay rate of Dalitz electrons from
the $\pi^0$ and $\eta$ mesons, and 
the result is given by $a = 4.9 \times 10^{-3}$ (see \ref{app2}).

The initial condition of electron spectrum at the top of the atmosphere
is the primary electron spectrum \cite{apj80} 
and that of the gamma-ray spectrum is the Galactic diffuse gamma rays, 
which is assumed to be zero. We have
\begin{eqnarray}
 \pi(E, 0) & = & 1.6\times10^{-4}( \frac{E}{100\ {\rm GeV}})^{-3.3} 
 \label{ele_ini} \\
 & & {\rm (m^2 \cdot sr \cdot sec \cdot GeV)^{-1}} \nonumber 
 \ , \\
 \gamma(E,0) & = & 0 \ . \nonumber
\end{eqnarray}
The eq.~({\ref{ele_ini}) gives the electron spectral index of $\alpha=2.3$.

%
The transport equations are solved using the above conditions.
The solution of electron spectrum  at the depth $t$ becomes
\begin{eqnarray}
   \pi(E, t) & = & \pi(E, 0)\zeta(t)  +  \gamma_{ex}(E,t)\xi(t)  \label{eq:el}
   \ , 
   \\[4mm]
   \zeta(t) & = & \frac{
                 [ ( \lambda_{1} + \sigma_{0} ) e^{\lambda_{1}t}-
                 ( \lambda_{2} + \sigma_{0} ) e^{\lambda_{2}t} ]  }
                 {\lambda_{1}-\lambda_{2}}
   \ , 
                 	\nonumber  \\
   \xi(t)   & = & \frac{B(\beta)}{\lambda_{1}-\lambda_{2}}
                 [ \frac{e^{(\lambda_{1}+\frac{1}{L})t}-1}{\lambda_{1}+1/L}-
                 \frac{e^{(\lambda_{2}+\frac{1}{L})t}-1}{\lambda_{2}+1/L} ] 
                 	\nonumber \\ 
			& &	+  \frac{a}{\lambda_{1}-\lambda_{2}} \times
					\nonumber \\
       		& &	[ \frac{(\lambda_{1}+\sigma_0)e^{(\lambda_{1}+1/L)t} - (\sigma_0 - 1/L)}{\lambda_{1}+1/L} 
         		 	\nonumber \\
		    & &  - \frac{(\lambda_{2}+\sigma_0)e^{(\lambda_{2}+1/L)t} - (\sigma_0 - 1/L)}{\lambda_{2}+1/L} 
         		 ]
    \ .
					\nonumber
\end{eqnarray}
The coefficient $\zeta(t)$ represents the energy loss rate of electrons 
by bremsstrahrung in the atmosphere. 
The term $\gamma_{ex}(E,t)\xi(t)$ represents the atmospheric electron spectrum.
The first term of $\xi(t)$ represents the pair production rate from gamma rays and 
the second term including $a$ is the contribution from Dalitz decay mode.
%
The atmospheric gamma-ray spectrum at the depth $t$~${\rm g/cm^2}$ is given by 
\begin{eqnarray}
 \gamma(E, t) & = & \pi(E, 0)\eta_{1}(t) + \gamma_{ex}(E,t)\eta_{2}(t)  \label{eq:ga} 
 \ , \\[4mm]
   \eta_{1}(t) & = & \frac{C(\alpha)}{\lambda_{1}-\lambda_{2}}
                     [ e^{\lambda_{1}t}-e^{\lambda_{2}t} ]
					\ , \nonumber \\
   \eta_{2}(t) & = & \frac{1}{\lambda_{1}-\lambda_{2}}[f(\lambda_{1}) - f(\lambda_{2}) ]
   					\ , \nonumber \\[2mm]
      & &  f(\lambda) =  \frac{1}{\lambda+1/L} \{ (\lambda+A(\beta) )e^{(\lambda+1/L)t}
                    \nonumber \\ 
      & &  \ \  -(A(\beta)-\frac{1}{L}) +  a \cdot C(\beta) ( e^{(\lambda+1/L)t} - 1 ) \}
      				\ , \nonumber
%
\end{eqnarray}
where $\eta_1$ represents the gamma-ray production rate from primary electrons and 
$\eta_2$ represents the attenuation rate of gamma rays by pair production.
The coefficient $\lambda_1(s)$, $\lambda_2(s)$, 
$A(s)$, $B(s)$, $C(s)$ and $\sigma_0$ with $s = \alpha, \ \beta$ used above equations are 
well known formulas and given in \ref{app1}.
%

The approximate expression of atmospheric electron spectrum becomes
\[
\xi(t) \cdot \gamma_{ex}(E, t) \sim 
( \frac{B(\beta)}{2} t^2 + a \cdot t ) \gamma_{ex} (E, t)
\ , 
\]
which corresponds to the estimate of atmospheric electron spectrum
in the Orth\&Buffington's paper \cite{orth}.

The production rate of gamma rays eq.~(\ref{ga_ex}) estimated from the eq.~(\ref{eq:ga}) 
and the eq.~(\ref{ga_ob}) is given by
\begin{eqnarray*}
\gamma_{ex}(E, t) & = & 1.11 \times 10^{-3}( E/100~{\rm GeV})^{-2.73} \exp(\frac{-t}{100~{\rm g/cm^2}})\\
& & {\rm (m^{2} \cdot s \cdot sr \cdot GeV \cdot r.l.)^{-1}} 
\ ,
\end{eqnarray*}
in which the radiation length of air is taken as 36.6~${\rm g/cm^2}$ \cite{pdb}.
%
\subsection{Solution with Zenith Angle Distribution}
As cosmic rays come in all directions, 
the intensities in the previous section should be integrated 
with the acceptance angle $\theta$ of each detector. 
At the depth $t$ , 
the electron differential spectrum $j_{ob}(E, t)$ and 
the gamma-ray spectrum $\gamma_{ob}(E, t)$
are derived as follows.
\begin{eqnarray}
j_{ob}(E, \theta, t) \ ( {\rm m^{2} \cdot sec \cdot GeV^{-1}}) \hspace{25mm} & & \nonumber \\
  				=   2\pi \int_{0}^{\theta} 
     				\pi(E, \frac{t}{\cos\theta})\cos\theta \sin\theta d\theta \hspace{10mm}& &
 \nonumber  \\
                =  \pi(E, 0)      \zeta(\alpha, \theta, t)
                    + \gamma_{ex}(E,t) \xi(\beta, \theta, t) & &
\label{e_ob}   \\[3mm]
\gamma_{ob}(E, \theta, t) \ ({\rm m^{2} \cdot sec \cdot GeV^{-1}} )\hspace{25mm} & & \nonumber \\
  				=   2\pi \int_{0}^{\theta}
   					\gamma(E, \frac{t}{\cos\theta}) \cos\theta \sin\theta d\theta \hspace{13.2mm}& & 
   	\nonumber \\
  				=  \pi(E, 0)      \eta_{1}(\alpha, \theta, t)
                     + \gamma_{ex}(E,t) \eta_{2}(\beta, \theta, t) & &  
\nonumber
\end{eqnarray}
The coefficients of each term are calculated from the following expressions, 
\begin{eqnarray*}
   \zeta(\alpha, \theta, t) & = &  
   							O_{1}(\alpha, \theta, t) + \sigma_{0} O_{2}(\alpha, \theta, t)  \\
   \xi(\beta, \theta, t) & = &  B(\beta)O_{3}(\beta, \theta, t) \\
						 &   &  + a \cdot \{ O_{2}(\beta, \theta, t)
                                 + \sigma_0 O_{3}(\beta, \theta, t) \}   \\
   \eta_{1}(\alpha, \theta, t) & = &  C(\alpha) O_{2}(\alpha, \theta, t) \\
   \eta_{2}(\beta, \theta, t)  & = & O_{2}(\beta, \theta, t) \\
                               &   & + (A(\beta)+a \cdot C(\beta)) O_{3}(\beta, \theta, t)
\end{eqnarray*}
and the series of $O_{l}(s, \theta, t)$ with $l=1,2,3$ are given by
\[
 O_{l}(s, \theta, t)  =  2\pi
                           \sum_{n=0}^{\infty} \Lambda_n(s) T_{n-1+l}(\theta,t)\\
\]
%
\[
     	 T_0  =  \frac{1-\cos^{2}\theta}{2}, \ \ \ 
         T_1  =  t (1 - \cos\theta ), 
\]
\[
         T_2  =  \frac{t^{2}}{2!}\log(\frac{1}{\cos\theta}), \ \ \ 
         T_3  =  \frac{t^{3}}{3!}(\frac{1}{\cos\theta}-1), \ \ \ 
\]
\[
         \cdots \ \ \
         T_n  =  \frac{t^{n}}{n!}\frac{1}{n-2}(\frac{1}{\cos^{n-2}\theta}-1) \ \ \ 
		 \cdots \\
\]
\[  \Lambda_{0} =  \frac{\lambda_1-\lambda_2}{\lambda_1-\lambda_2}, \ \cdots, 
\ \Lambda_{n} = \frac{\lambda_1^{n+1}-\lambda_2^{n+1}}{\lambda_1-\lambda_2} +
  					 \Lambda_{n-1} \frac{-1}{L}, \ \cdots 
\]
%
\section{Results and Discussions}
The analytic solutions obtained here are compared with the 
results of Monte Carlo simulation(MC).
We use the COSMOS simulation code \cite{cosmos} in which 
the Dpmjet3 \cite{dpm} is adopted as the hadronic interaction model and
the BESS-TeV \cite{bess1}\cite{bess2} and RUNJOB \cite{runjob} data 
for the input proton, helium  and CNO spectrum at the top of atmosphere.
We have confirmed the consistency of both methods for the cross section
using the fundamental processes as shown in \ref{app1}.
%
In addition, we consider atmospheric electrons directly originating 
from $K$ and $\pi^{\pm}$ mesons, which mainly contribute below 100GeV.
The decay modes of $\pi^{\pm} \rightarrow \mu^{\pm}\nu(\bar{\nu}) \rightarrow e^{\pm}\nu\bar{\nu}$, 
$K^0_L \rightarrow \pi^{\pm} e^{\mp} \nu$, $K^{\pm} \rightarrow \pi^0 e^{\pm} \nu $
and so on, are calculated using the recent accelerator data \cite{alice},
and those contributions are treated as the ratio to electrons produced by gamma rays.
%
\subsection{Atmospheric Electron and Gamma-ray Spectrum}
%
 \begin{figure}
 \begin{center}
   \includegraphics*[width=10cm]{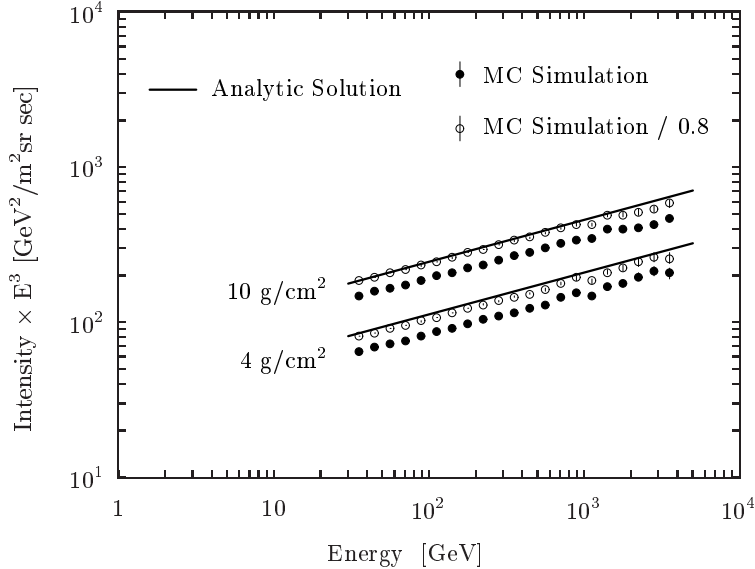}
 \end{center}
 \caption{Atmospheric gamma-ray spectra $J_g(E)$ of eq.~(\ref{eq:ga}) 
 are compared with the MC results. 
  The solution at 4~${\rm g/cm^2}$ is equal to eq.~(\ref{ga_ob}) multiplied by $E^3$.
The MC data divided by 0.8 are also shown and good agreement with
the analytic solution. 
}
 \label{fig1}
 \end{figure}
The analytic solution of atmospheric gamma-ray spectrum given by 
\begin{equation}
J_g(E, t) = \gamma_{ex}(E,t)\eta_{2}(t)
\label{gamma}
\end{equation}
in eq.~(\ref{eq:ga}) and is shown in Figure~\ref{fig1} 
with the MC results at the depth of 4,~10~${\rm g/cm^2}$.
Figure~\ref{fig1} shows that the gamma-ray spectrum of MC is 20\% lower than the analytic solution, 
which has been indicated in our previous paper \cite{yoshida}:
the proton spectrum deconvolved from our observed gamma-ray spectrum 
assuming the nuclear interaction model of Dpmjet3
is 20\% higher than observed primary proton data.
%
Thus the discrepancy of gamma-ray flux in this paper is consistent with the difference 
between deconvolved and observed proton flux 
as described in the previous paper \cite{yoshida}.

 \begin{figure}
 \begin{center}
  \includegraphics*[width=10cm]{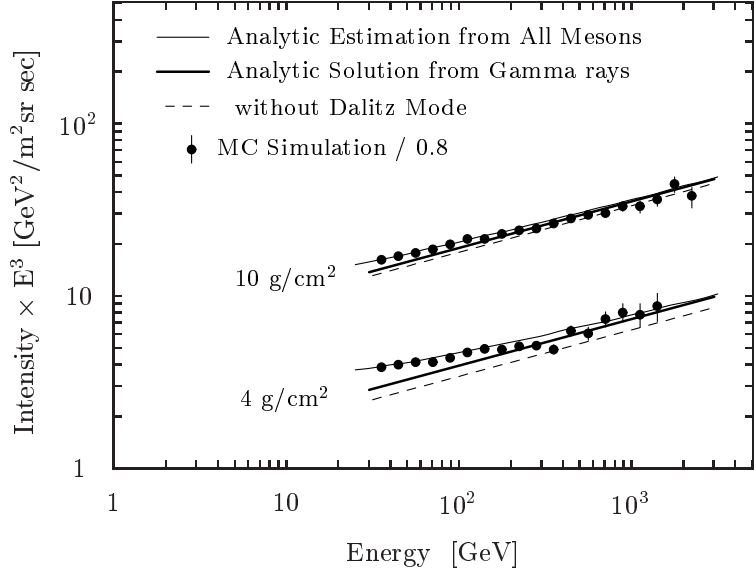}
 \end{center}
 \caption{Atmospheric electron spectra are shown 
 at the depth 4, 10~${\rm g/cm^2}$. The analytic estimation from all mesons 
 (eq.~(\ref{sec_t})) and the analytic solution $\xi(t) \gamma_{ex}(E,t)$ 
 estimated from gamma-ray spectrum  with and without Dalitz decay mode 
 are shown with MC results divided by 0.80.
}
 \label{fig2}
 \end{figure}

The analytic estimation of atmospheric electron spectrum at the depth $t$ is given by
\begin{equation}
j_{sec}(E, t) = (1 + C) \xi(t) \gamma_{ex}(E,t)
\label{sec_t}
\end{equation}
as shown in eq.~(\ref{eq:el}). The fraction $C$ represents the contribution 
from direct productions from $K$ and $\pi^{\pm}$ mesons and is shown as 
the increase of the thin line(all contributions) to the thick line(from gamma rays) 
in  Figure~\ref{fig3}.
In the fraction $C$, the $K^0_L$ mode are mostly dominant as indicated by 
Orth \& Buffington \cite{orth}. The contribution from $K$ mesons estimated 
from the recent accelerator data of $K/\pi$ ratio \cite{alice} is a factor of two 
smaller than their result.

The atmospheric electron intensity is approximately proportional to $t^2$.
The comparison with MC is shown in Figure~\ref{fig2}, in which 
the spectrum of MC is divided by 0.80 to correct the discrepancy of gamma-ray spectrum.
%
In the energy range above several hundred GeV, 
we notice that the Dalitz decay mode contributes more than 10~\% to the ratio
at high altitude above 4~${\rm g/cm^2}$, and 
the estimated spectrum is good agreement with the corrected MC simulation
in all energy range.
The atmospheric electrons become a larger fraction of 
the observed electrons in the higher energy region 
because of the harder spectrum than the primary electron one.
%
 \begin{figure}
 \begin{center}
  \includegraphics*[width=10cm]{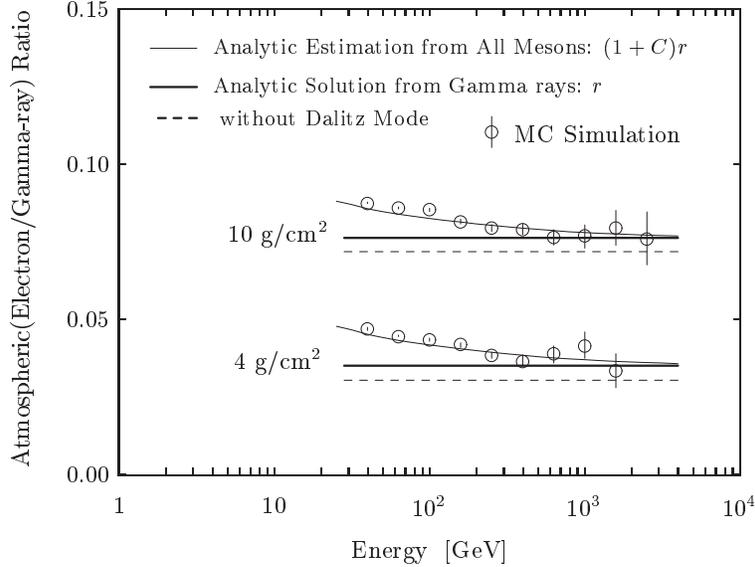}
 \end{center}
\caption{The ratio of atmospheric electrons to gamma-rays.  The analytic estimation
from all mesons and the analytic solution from gamma-ray spectrum
with and without Dalitz mode are compared with Monte Carlo simulation.
The ratio $r$ represents $\xi(t)/\eta_2(t)$ and 
the fraction $C$ represents the contribution 
from direct productions from $K$ and $\pi^{\pm}$ mesons.
}
  \label{fig3}
 \end{figure}

We investigate the consistency between the analytic estimation and MC 
by comparing the ratio of atmospheric electrons to gamma rays,
which is able to exclude the discrepancy of gamma ray spectrum.
The ratios compared at the depth $t=$ 4, 10~${\rm g/cm^2}$ are shown in Figure~\ref{fig3}, 
which shows good agreement between the analytic estimation and MC simulation.
The Dalitz contribution has constant value and does not vary with the depth $t$,
since Dalitz electrons approximately increase  in proportional to 
gamma rays and $t$,
while electrons produced by gamma rays are proportional to $t^2$.
It means that the Dalitz mode plays a more important role at the higher altitude
above several hundreds GeV.
%
On the other hand, below 100GeV the direct productions from $K$ and $\pi^{\pm}$ mesons
become larger than that from the Dalitz mode.
%
\subsection{Observation and Correction}
 \begin{figure}
 \begin{center}
  \includegraphics*[width=10cm]{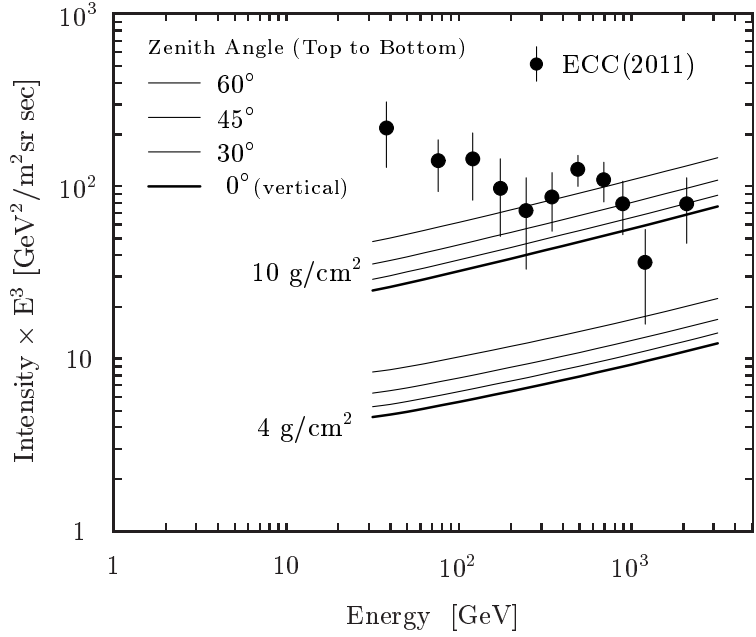}
 \end{center}
\caption{ 
The correction term of atmospheric electrons of eq.~(\ref{eq:sec_c})
with acceptance angle of 30$^{\circ}$, 45$^{\circ}$, 60$^{\circ}$ and vertical
for detectors at 4, 10~${\rm g/cm^2}$, 
and compared with the result of primary electron spectrum.
}
\label{fig4}
 \end{figure}
%
%
%
 \begin{figure}
 \begin{center}
  \includegraphics*[width=10cm]{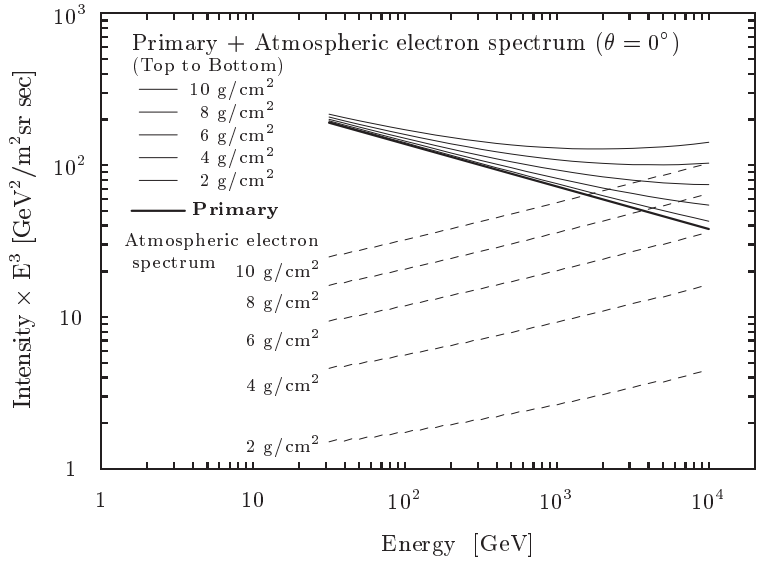}
 \end{center}
\caption{ The primary plus atmospheric electron spectrum (thin line)
 at each depth above 10~${\rm g/cm^2}$ including the extrapolation above 3~TeV.
The primary electron spectrum (thick line) with the simple power law is estimated form the 
recent result below 3~TeV \cite{apj11}.
The atmospheric electron spectra (dashed line) are given by eq.~(\ref{eq:sec_c}).
}
\label{fig5}
 \end{figure}
The primary electron spectrum is obtained from 
\begin{equation}
 \pi(E, 0)  =  \frac{j_{ob}(E, \theta, t)}{\zeta(\alpha, \theta, t)}
           -  \gamma_{ex}(E,t)\frac{\xi(\beta, \theta, t)}{\zeta(\alpha, \theta, t)}
\ \ {\rm (m^{2} \cdot sr \cdot sec \cdot GeV)^{-1} } \ , 
\label{eq:primaryflux}
\end{equation}
using the eq.~(\ref{e_ob}), in which $1/\zeta$ is the correction term of
the energy reduction by bremsstrahrung.
The second term is the correction term of atmospheric electrons 
which are observed at the depth $t$ within zenith angle $\theta$ and 
corrected by energy reduction to the top of the atmosphere.
The term $\xi/\zeta$ is approximately proportional to $t^2$.
The eq.~(\ref{eq:primaryflux}) indicates that 
all the observed electrons are firstly corrected by bremsstrahrung loss energy, 
and then, the atmospheric electrons are statistically subtracted.
%

Finally the atmospheric electron spectrum corrected at the top of the atmosphere 
is given by 
\begin{equation}
 J_{sec}(E, t)  = (1 + C) \gamma_{ex}(E,t)\frac{\xi(\beta, \theta, t)}{\zeta(\alpha, \theta, t)}
\ \ {\rm (m^{2} \cdot sr \cdot sec \cdot GeV)^{-1} } 
\ .
\label{eq:sec_c}
\end{equation}
Figure~\ref{fig4} indicates the correction term of 
atmospheric electron spectra of eq.~(\ref{eq:sec_c}) 
for the different acceptance angle of each detector at 4, 10~${\rm g/cm^2}$, 
compared with the primary electron spectral data.
%
The fraction $C$ practically has no influence on the correction of the primary electron spectrum.
%
%
%
%
Further in Figure~\ref{fig5}, we show the primary electron spectrum, the atmospheric electron spectra
of eq.~(\ref{eq:sec_c}) and total spectra at each depth above 10~${\rm g/cm^2}$.
%
%
It is shown that the balloon experiment for electrons above 1~TeV requires
the observation at a high altitude, since correction terms 
dominate the data at 10~${\rm g/cm^2}$.
%
\section{Conclusions}
In the balloon observations of cosmic-ray electrons, 
the atmospheric electrons are inevitably included in the observed spectrum
and necessary to be subtracted to obtain the primary electron spectrum.
%
The atmospheric electrons are mostly produced by the atmospheric gamma-ray conversion, 
and above several hundred GeV, around 10\% of them at the depth of 4~${\rm g/cm^2}$ 
are produced by the Dalitz decay mode.
We have treated these processes analytically and obtained accurate solutions.
The result shows that the atmospheric electron spectrum obtained by MC 
is 20\% lower than that of the analytic estimation. 
The discrepancy comes from the difference of atmospheric
gamma-ray spectrum, 
since the ratio of atmospheric electrons to gamma rays
shows a good agreement.
The difference of the gamma-ray spectrum comes from 
the uncertainties of the primary cosmic-ray hadron spectra 
and the nuclear interaction models used in the MC calculations, 
together with the statistical error of 10~\% 
included in the observed gamma-ray spectrum of eq.~({\ref{ga_ob}).
The Dalitz mode gives a constant contribution to the atmospheric 
electron$/$gamma-ray ratio and 
 becomes particularly important at the higher altitude above 4~${\rm g/cm^2}$.
On the other hand, in the lower energy range below 100~GeV 
the direct contributions of electrons from  $K$ and $\pi^{\pm}$ mesons 
become larger than the Dalitz contribution, although atmospheric electrons 
below 100 GeV have almost no influence on the correction of the primary electron spectum.
%
%

The astrophysical significance to search for the cosmic ray sources 
by primary electrons become more important at higher energy regions, 
however, the contribution of the background atmospheric electrons 
increases seriously in the balloon observations in TeV region.
Thus the observation of TeV electrons requires the higher altitude,
and more precise estimates of the atmospheric electrons 
are required than that of several hundred GeV region.
Although these conditions 
are getting more difficult at higher energy region,
the experiments in TeV region are the most important 
to search for cosmic-ray sources \cite{apj04}.
%
\section{Acknowledgment}
We sincerely thank 
Prof.~J.~Arafune for his helpful discussions on the contribution of 
the Dalitz pair to the atmospheric electrons.
We also thank Dr.~M.~Honda for his valuable comments on the nuclear interaction data.
We are grateful to Prof. R. J. Wilkes for his careful reading
and useful comments of the manuscript.

%

%
\newpage
\appendix
\section{Fundamental Parameters}	\label{app1}
We calculate the fundamental process and compare the result with that of MC
to confirm the consistency of cross section.
When gamma rays or electrons pass through the thickness $t$ of the air,
the number of electrons $\pi(E, t)dE$ and gamma rays $\gamma(E, t)dE$ 
with energies between $E$ and $E+dE$ satisfy the transport equations as:
\begin{eqnarray*}
  \frac{\partial \pi (E, t)}{\partial t} & = & 
   -A' \pi (E, t) + B' \gamma (E, t)            \\
  \frac{\partial \gamma (E, t)}{\partial t} & = & 
   C' \pi (E, t) - \sigma_{0}\gamma(E, t) 
\end{eqnarray*}
The physical meaning of each parameter $A'$, $B$', $C'$ and $\sigma_0$
is explained in the section~\ref{sec:sol} in this text.
The functional forms
of the operators $A'$, $B'$, $C'$, $\sigma_0$ and $\lambda_{1, 2}$ used in the equations 
are given in several references (e.g. \cite{HB}) as 
\begin{eqnarray*}
  A(s) & = & 1.3603 \frac{d}{ds}\ln\Gamma(s+2) - \frac{1}{(s+1)(s+2)} - 0.07513
   \ , \\
  B(s) & = & 2 ( \frac{1}{s+1} - \frac{1.3603}{(s+2)(s+3)} )
   \ , \\
  C(s) & = & \frac{1}{s+2} + \frac{1.3603}{s(s+1)}
   \ , \\
  \sigma_{0} & = & 0.7733 \ , \\
  \lambda_{1, 2}(s) & = & \frac{1}{2}[ -(A(s)+\sigma_0) \ \pm \  \{ (A(s)+\sigma_0)^2 \\
 			  &   & \hspace{10mm} - 4(A(s)\sigma_0-B(s)C(s))\}^{\frac{1}{2}} ]  
\ ,
\end{eqnarray*}
where $s+1$ corresponds to a power-law index of the solutions.
The numerical values of $A(s)$, $B(s)$ and $C(s)$ used in this paper 
are listed as 
\[ A(2.30)= 1.6743\ , \ B(2.30)= 0.4867 \ , \ C(2.30)= 0.4118 \ , \]
\[ A(1.73)= 1.4270\ , \ B(1.73)= 0.5784 \ , \ C(1.73)= 0.5561 \ . \]

We calculate the gamma ray conversion process and 
confirm the agreement between analytical method and MC.
If the initial condition is given by 
\begin{eqnarray*}
 \pi(E, 0) & = & 0 \ , \\
 \gamma(E,0) & = &  k_g ( \frac{E}{100\ {\rm GeV}} )^{-\beta-1} 
\end{eqnarray*}
with $\beta=1.73$, the solution of produced electron and gamma ray spectrum at the depth $t$ becomes 
\begin{eqnarray*}
	\pi(E, t) & = & \frac{\gamma(E,0)}{\lambda_{1}-\lambda_{2}} B(\beta)
                     [ e^{\lambda_{1}t} -  e^{\lambda_{2}t} ]  , \\
	\gamma(E, t) & = & \frac{\gamma(E,0)}{\lambda_{1}-\lambda_{2}}
                      [    A(\beta)(e^{\lambda_{1}t}-e^{\lambda_{2}t} ) \\
                 &   &  \hspace{15mm}   +  \lambda_1 e^{\lambda_{1}t} - \lambda_2 e^{\lambda_{2}t} ]
\ .
\end{eqnarray*}
In Table~\ref{table2}, 
the number of electrons $\pi(E, t)$ produced by gamma ray conversion are 
shown at the depths of 0.4~${\rm g/cm^2}$ and 4.0~${\rm g/cm^2}$
and compared with the results of MC.
These values directly represent the cross section of pair production 
and one can find that the cross section agrees with each other within $\pm 0.5$~\%.
%
%
\begin{table}
\begin{center}
\caption{
The number of Electrons produced by gamma rays at 100~GeV, which is
divided by the initial number of $k_g$.
}
\begin{tabular}{c|cc} \hline
 Depth ($t$)  &   0.4~${\rm g/cm^2}$    &  4.0~${\rm g/cm^2}$   
\\ \hline
Analytic Solution & 6.229$\times 10^{-3}$ & 5.596$\times 10^{-2}$  \\
MC  & (6.25 $\pm$ 0.04)$\times 10^{-3}$ &  (5.61 $\pm$ 0.04)$\times 10^{-2}$  \\ \hline
\end{tabular}
\label{table2}
\end{center}
\end{table}
%
\section{Dalitz electron contribution}  \label{app2}
We include the contribution of Dalitz electrons 
represented by a fraction of gamma-ray production rate (eq~(\ref{ratio_a})).
The coefficient $a$ is estimated from the sum of Dalitz electrons of 
$\pi^0$, $\eta$ and $K^0$ decays.

In the $\pi^0$ decay, 
the branching ratio of the first mode is given  by
$B_m \equiv B(\pi^0 \rightarrow \gamma\gamma) = (98.798 \pm 0.032)\%$ and 
that of the second mode is
$B_D \equiv B(\pi^0 \rightarrow e^+e^-\gamma) = (1.198 \pm 0.033)\%$ \cite{pdb}.
We require the the ratio of Dalitz electrons to gamma rays 
from the $\pi^0$ meson spectrum $\propto E^{-\beta-1}$ ($\beta = 1.73$), and it
becomes  
$B_D/(2B_m)/(2/(\beta+1))= 4.4 \times 10^{-3}$, if the distribution of pair electrons is uniform.
%
Precise estimate gives a somewhat different distribution of the pair production 
with two peaks, so that the value is slightly corrected and becomes  $4.6 \times 10^{-3}$.

The ratio of Dalitz electrons to gamma rays for each decay
is listed in Table~\ref{table3}.
The major contribution of Dalitz electrons comes from $\pi^{0}$ decay.
The $\eta$ decay gives the small contribution, and the $K^0$ decay can be neglected.
%
The net fraction of Dalitz decay electrons to the observed gamma rays 
are given by $4.9 \times 10^{-3}$.
%
%
\begin{table}
\begin{center}
\caption{Dalitz contributions of $\pi^{0}$, $\eta$ and $K^0$ decays}
\begin{tabular}{l|rrr} \hline
  & $\pi^{0}$	& $\eta$  & $K^0$ \\ \hline \hline
Gamma-ray ($\gamma\gamma$) mode
 &	98.8~\% 	&	39.3~\%	& 98.8~\% \\
Dalitz ($e^+e^-\gamma$) mode
 &  1.2~\% 	&  0.68~\%  	& 1.2~\% \\
%
Dalitz electron $/$ $\gamma$-ray ($r$)
& $4.6\times 10^{-3}$ 	& $6.6\times 10^{-3}$ 		& $4.6 \times 10^{-3}$ \\
Fraction of $\gamma$-ray production ($p$)
& 1/1.19 				& 0.16/1.19   		& 0.03/1.19 \\
%
Ratio in upper atmos. ($= r \times p $)
& $3.9\times 10^{-3}$ 	& $0.9\times 10^{-3}$ 	& $0.1\times 10^{-3}$ \\ 
\hline
\end{tabular}
\label{table3}
\end{center}
\end{table}
%
\end{document}